\documentclass[twocolumn,showpacs,preprintnumbers,amsmath,amssymb,times]{revtex4}
\usepackage{graphicx}
\usepackage{dcolumn}
\usepackage{bm}

\begin{document}
\title{The effect of bandwidth in scale-free network traffic}

\author{Mao-Bin Hu$^1$}\email{humaobin@ustc.edu.cn}
\author{Wen-Xu Wang$^2$}
\author{Rui Jiang$^1$}
\author{Qing-Song Wu$^1$}\email{qswu@ustc.edu.cn}
\author{Yong-Hong Wu$^3$}

\affiliation{
$^{1}$School of Engineering Science, University of
Science and Technology of China, Hefei 230026, P.R.C \\
$^{2}$Nonlinear Science Center and Department of
Modern Physics, University of Science and Technology of China,
Hefei 230026, P.R.C\\
$^3$Department of Mathematics and Statistics, Curtin University of Technology,
Perth WA6845, Australia}

\date{\today}

\begin{abstract}
We model information traffic on scale-free networks by introducing 
the bandwidth as the delivering ability of links. 
We focus on the effects of bandwidth on the packet delivering ability 
of the traffic system to better understand traffic dynamic in 
real network systems. 
Such ability can be measured by a phase transition from free flow 
to congestion.
Two cases of node capacity $C$ are considered, i.e., 
$C=constant$ and $C$ is proportional to the node's degree.
We figured out the decrease of the handling ability of 
the system together with the movement of the optimal local 
routing coefficient $\alpha_c$, induced by the restriction 
of bandwidth. 
Interestingly, for low bandwidth, the same optimal value of 
$\alpha_c$ emerges for both cases of node capacity. 
We investigate the number of packets of each node 
in the free flow state and provide analytical explanations
for the optimal value of $\alpha_c$.
Average packets traveling time is also studied. 
Our study may be useful for evaluating the overall efficiency of 
networked traffic systems, and for allevating traffic jam in such 
systems.
\end{abstract}


\maketitle

\section{Introduction}
Since the pioneering work of Watts and Strogatz \cite{WS}, and 
Barab\'{a}si and Albert \cite{BA}, complex networks theory 
have attracted growing interest among physics community.
Complex networks can describe many natural, social and technical 
systems in which lots of entities or people are connected by 
physical links or some abstract relations
\cite{BA2,BA3,Newman,Newman2,Boccaletti,Humaobin,WuJJ}. 
Due to the importance of large communication networks such as 
the Internet, WWW, power grid and transportation systems 
with scale-free properties in modern society,
the traffic of information flow have drawn more and 
more attention.
Ensuring free traffic flow on these networks is of great 
significance and research interest
\cite{Moreno,Sole,Arenas,Tadic,Zhao,Mukherjee,Guimera,Guimera2,
Goh,YanGang,Wang,Wang2,Yin,deMenezes,deMenezes2}.

As pointed out by Newman, the ultimate goal of studying complex 
networks is to understand how the network effects influence many 
kinds of dynamical processes taking place upon networks 
\cite{Newman}.
Recent works have proposed some models to mimic the traffic 
on complex networks by introducing packets generating rate 
and the routing of packets 
\cite{Sole,Arenas,Tadic,Zhao,Mukherjee,Guimera,Guimera2}.
These kind of models can define the capacity of networks 
by the critical generating rate at which a phase transition 
from free flow state to congested state occurs.
The free flow state corresponds to the number of created 
and delivered packet 
are balanced, and the jammed state 
corresponds to the packets accumulate on the network. 

Many recent studies have focused on two aspects to control 
the congestion and improve the efficiency of transportation: 
modifying underlying network structures or developing 
better route searching strategies in a large network 
\cite{Kleinberg}. 
Due to the high cost of changing the infrastructure, 
the latter is comparatively preferable.
In this light, some models have been proposed to forward packets 
following the shortest path \cite{Goh}, the efficient path 
\cite{YanGang}, the nearest-neighbor and next-nearest-neightbor 
searching strategy \cite{Wang,Wang2,Yin}, 
the local static information \cite{Wang}, or the integration of local 
static and dynamic information \cite{Wang2}. 
In view of the difficulty of knowing the whole topology of many 
large and rapidly growing communication systems, the local routing 
strategy attracts more attention because the local static topology 
information can be easily acquired and stored in each router.

However, previous studies usually neglect the bandwidth of
the links, i.e., the maximum capacity of each link for delivering 
packets.
Obviously, in real systems, the capability of links are limited 
and variates from link to link and in most cases, these restrictions 
could be very important in triggering congestion in the traffic 
system.
Therefore, since the analysis on the effects of the link bandwidth 
restrictions on traffic efficiency is still missing, 
we study the traffic dynamics in which the bandwidth is taken 
into account, based on the local routing strategy.

The paper is organized as follows. 
In section II, the traffic model is introduced. 
In section III, the simulation results are presented and discussed. 
The summary is given in section IV.

\section{The Traffic Model}
To generate the underlying traffic network, our simulation starts  with 
the most general Barab\'{a}si-Albert 
scale-free network model 
which is in good accordance with real observations of 
communication networks \cite{BA2}. 
Driven by the growth and preferential attachment mechanisms, 
it can generate power-law degree distribution 
$P(k) \sim k^{-\gamma}$, where $\gamma =3$.
In this model, starting from $m_0$ fully connected nodes, a new 
node with $m$ links is added to the existing graph at each time 
step according to the preferential attachment, i.e., 
the probability $\Pi_i$ of being connected to the existing node
$i$ is proportional to the degree $k_i$ of the node, 
$\Pi_i={k_i \over \Sigma_j k_j}$, where the sum runs over all existing
nodes. 

Then we model the traffic of packets on the given network. 
We treat all nodes as both hosts and routers for generating and 
delivering packets and assume that each node can deliver at most 
$C$ packets per step towards their destinations.	
The capacity of each link is restricted by Bandwidth($B$), 
i.e., each link can handle at most $B$ packets from each end 
per time step. 
Motivated by the previous local routing models \cite{Wang,Wang2},
the system evolves in parallel according to the following rules:

1. Add New Packets - Packets are added with a given rate $R$ 
(number of packets per time step) at randomly selected nodes 
and each packet is given a random destination.

2. Deliver Packets - Each node performs a local search among its 
neighbors. 
If a packet's destination is found in its nearest neighborhood, 
it will be delivered directly to its target and then removed 
from the system. 
Otherwise, its will be delivered to a neighboring node $n$ 
with preferential probability:
\begin{equation}
P_n={k^{\alpha}_n \over \Sigma_i k^{\alpha}_i},
\end{equation}
where the sum runs over the neighboring nodes, and $\alpha$ is a 
tunable parameter. 
FIFO (first-in-first-out) queuing discipline is applied at each node.

In the simulation, we investigate the network performance 
with node capacity $C=10$ or $C=k$ correponding to the degree of 
the node, and with link bandwidth $B=5,3,1$ and $1 \leq B \leq 5$ 
for each case. 

To characterize the system's overall capacity, we investigate 
the order parameter:
\begin{equation}
\eta(R)=\lim_{t \rightarrow \infty} { \langle \Delta N_p \rangle 
\over R \Delta t},
\end{equation}
where $N_p(t)$ is the number of packets on the network at time $t$. 
$\Delta N_p = N_p(t+\Delta t)-N_p(t)$ with $\langle ... \rangle$ 
takes average over time windows of width $\Delta t$. 
Obviously, $\eta(R)=0$ corresponds to the cases of free flow state 
where the balance between added and removed packets can be achieved.
As $R$ increases, there is a critical $R_c$ at which $\eta$ 
suddenly increase from zero to nonzero, which indicates the 
phase transiton from free flow to congestion that the packets 
begin to accumulate on the network. 
Hence, the system's capacity can be measured by the value of 
$R_c$ below which the system can maintain its efficient functioning.

\section{Simulation Results and Discussions}

\begin{figure}
\scalebox{0.9}[0.9]{\includegraphics{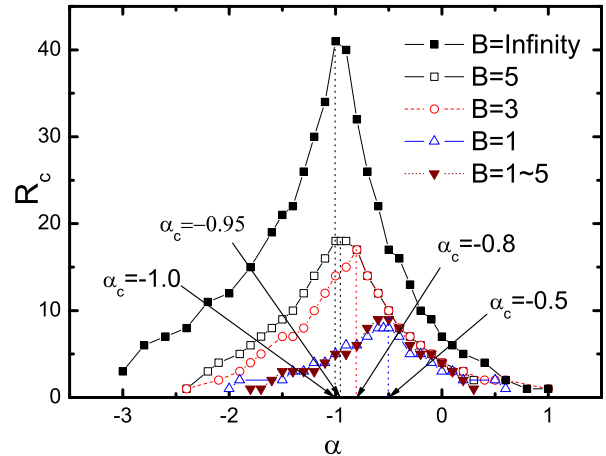}}
\caption{\label{Fig1}  (color online). 
The network capacity $R_c$ against $\alpha$ with network parameter 
$N=1000$, $m_0=m=3$, constant node delivering ability $C=10$, and 
different bandwidth $B$ cases. 
The data are obtained by averaging $R_c$ over 10 network realizations.}
\end{figure}

In the special case of $C=10$ and $B \geq 10$, the maximum network 
capacity is $R_c \approx 40$, which is achieved at the optimal 
value $\alpha_c=-1.0$ \cite{Wang}.  
This can be explained as the average number of packets on nodes 
does not depend on degree $k$ at $\alpha_c=-1.0$ and thus leads 
to a fact that no congestion occurs earlier on some nodes with 
particular degree than on others .

Then we study the effect of bandwidth on the network capacity 
in the case of fixed node capacity $C=10$ with constant bandwidth
$B=5,3,1$ for each link or $B$ is a random integer selected 
in the range from $1$ to $5$ for the links. 
The constant $B$ case corresponds to a uniform bandwidth system, 
and the random $B$ case corresponds to a system with different 
bandwidth for each link. 
Fig.\ref{Fig1} compares the network capacity $R_c$ for the cases. 
One can see that at a given $\alpha$, the network capacity decreases 
with small $B$ value. 
This is easy to be understood because the bandwidth prohibit the 
free flow of packets from one node to the other node thus 
decreases the network capacity. 

Furthermore, the optimal value of $\alpha_c$ corresponding to the 
maximum capacity increases from $-1.0$ to $-0.95$ for $B=5$, 
$-0.8$ for $B=3$, and $-0.5$ for both $B=1$ and $1\leq B\leq 5$. 
This can be explained as follows. 
Let $n_i(t)$ denotes the number of packets of node $i$ 
at time $t$.
In the case of homogeneously generated sources and destinations
for the packets, the number of packets generated and removed at 
node $i$ are balanced. 
Considering the contribution of received and delivered packets 
of node $i$ to the change of $n_i(t)$, the evolution of $n_i(t)$
in the free flow state can be written as 

\begin{equation}\label{eq3}
{dn_i(t) \over dt}=-n_{out}+n_{in},
\end{equation}
where $n_{out}$ denotes the number of packet delivered from node 
$i$ to its neighbouring nodes, and $n_{in}$ denotes the number of 
received packets. 
From Eq.\ref{eq3}, in the case of $B \geq C$, Wang et al. show 
that $n(k) \sim k^{1+\alpha}$ \cite{Wang}. 
Therefore, when $\alpha=-1.0$, the average number of packets 
on nodes is independent of degree $k$ and thus there will not 
be some nodes that are more easy to jam, so that the maximum 
network capacity is achieved. 
However, $\alpha>-1.0$ means that there are more packets on 
the hub nodes (with greater degree $k$).
Considering the restriction of $B<C$, since the 
hub nodes have more links thus have more total bandwidth, 
$\alpha>-1.0$ is better to fully use the bandwidth of the 
hub nodes and thus enhance the system's capacity. 

To better understand why $\alpha >-1.0$ is the optimal 
choise, we investigate the number of received packets 
of node $i$

\begin{equation}\label{eq4}
n_{in}(i)=\sum_{j=1}^N A_{ij} n_j P_i 
= \sum_{j=1}^N A_{ij} n_j {k_i^\alpha \over 
\sum_{l=1}^N A_{jl} k_l^\alpha},
\end{equation}
where the sum run over all the nodes of the network and 
$A_{ij}$ is the element of the adjacency matrix. 
Considering that the assortative mixing of BA network is zero,
i.e., the average neighbors¡¯ degree of each node is the same, 
therefore we can get

\begin{equation}\label{eq5}
\sum_{l=1}^N A_{jl}k_l^\alpha 
= \sum_{l=1}^N A_{jl} W
= k_j W,
\end{equation}
where $W$ is a constant. In order to keep in the 
free flow state, one can easily conclude from Eq.\ref{eq3}
that $n_{out} \geq n_{in}$ should be satisfied. 
For high-degree nodes, $n_{out}$ is mainly constrained by 
two limits: $n_{out} \approx Bk_i$ and $n_{out} \approx C$.
Considering $n_{out} \approx Bk_i$, and inserting 
Eq.\ref{eq5} into Eq.\ref{eq4}, we can get 

\begin{equation}\label{eq6}
Bk_i \geq \sum_{j=1}^N A_{ij} n_j {k_i^\alpha \over k_j W}.
\end{equation}

Since $C$ is a constant, higher degree nodes are more easily 
congested than those low degree nodes.
We consider the case that $i$ is a high degree node, for 
BA scale-free network, most of neighbors of $i$ are low 
degree nodes. 
For small $B$, $n_{out}$ of low degree 
nodes are mostly restricted by the bandwidth. 
Hence, we assume a linear relationship for low-degree 
nodes as

\begin{equation}\label{eq7}
n_j =B k_j
\end{equation}

Insert Eq.\ref{eq7} into Eq.\ref{eq6}, we get

\begin{equation}\label{eq8}
\alpha \leq {\log W \over \log k_i}.
\end{equation}

In the limit of very large network, 
$N \rightarrow \infty$, $k_i \rightarrow \infty$ and 
thus the right hand side of Eq.\ref{eq8} approaches 
zero, and so that the optimal $\alpha$ should be 
smaller than zero.

Considering $n_{out} \approx C$, we can get 

\begin{equation}\label{eq9}
C \geq \sum_{j=1}^N A_{ij} n_j {k_i^\alpha \over k_j W} 
\end{equation}

Inserting Eq.\ref{eq7} to Eq.\ref{eq9}, we obtain 

\begin{equation}\label{eq10}
\alpha \leq {\log {CW\over B} \over \log k_i} -1.
\end{equation}

For low-degree nodes, $n_{out}$ is mainly constrained 
by $Bk_i$, i.e., $n_{out} \approx Bk_i$. 
And for BA scale-free network, most of neighbors of 
low-degree nodes are high-degree nodes, for which 
we can take $n_j \approx C$. 
Thus we can get

\begin{equation}\label{eq11}
Bk_i \geq \sum_{j=1}^N A_{ij} C {k_i^\alpha \over k_j W}.
\end{equation}

So we obtain

\begin{equation}\label{eq12}
\alpha \leq {\log{BW \over W'C} \over \log k_i},
\end{equation}
where the constant $W'= \sum_{j=1}^N A_{ij} {1 \over k_j}$. 
To continue, we use Eq.\ref{eq10}, which is the 
minimum of Eq.\ref{eq8},\ref{eq10} and \ref{eq12}.
One can see from Eq.\ref{eq10} that the optimal $\alpha$ 
should locate between $-1.0$ and $0.0$, and that 
when $B$ decrease from infinity to 1, $\alpha_c$ will 
be more close to zero.
For our simulation parameters, we can get 
$W \approx 0.4$ when $\alpha=-0.5$ (Eq.\ref{eq5})
and $k_{max} \approx 100$, 
thus $\alpha_c \leq -0.7$ for the case of $B=1$. 
It is quite close to our simulation result 
(see Fig.\ref{Fig1}).

\begin{figure}
\scalebox{0.9}[0.9]{\includegraphics{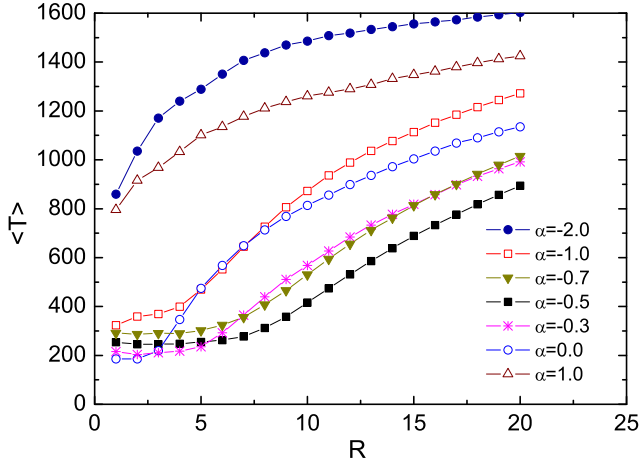}}
\caption{\label{Fig2}  (color online). 
The variation of packets average traveling time 
$\langle T \rangle$ versus $R$ with different value 
of $\alpha$ fixed. 
Other network parameters are $N=1000$, $m_0=m=3$,
$C=10$ and $B=1$. }
\end{figure}

Then we simulate the packets' average traveling time 
which is also important for measuring the system's 
efficiency. 
In Fig.\ref{Fig2}, we show the average traveling time 
$\langle T \rangle$ against $R$ with different $\alpha$.
One can see for $-1.0 \leq \alpha \leq 0.0$, 
$\langle T \rangle$ remain as a relatively small value 
when $R \leq R_c(\alpha)$. 
When $R$ increases beyond $R_c$, $\langle T \rangle$ will 
increase very rapidly, implying that the system is jammed.
When $\alpha=-0.5$, the optimal average traveling time 
is obtained, whereas $\langle T \rangle$ increase much 
more rapidly when $\alpha$ deviates from $-0.5$.
Thus $\alpha_c=-0.5$ is the best choice.
This is consistent with the above analysis that a 
maximum $R_c$ occurs when $\alpha_c=-0.5$. 

\begin{figure}
\scalebox{0.9}[0.9]{\includegraphics{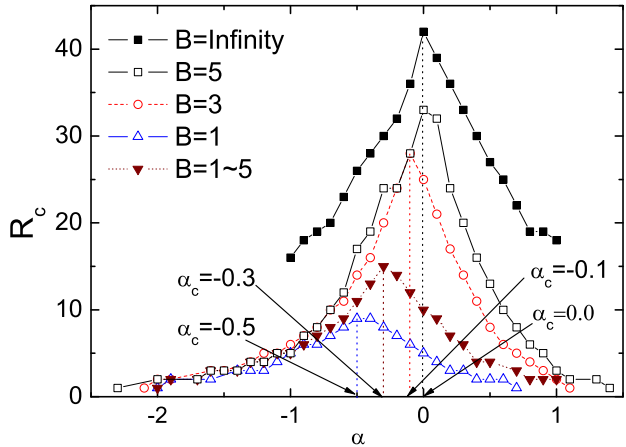}}
\caption{\label{Fig3}  (color online). 
The network capacity $R_c$ against $\alpha$ in the case of 
node delivering ability proportional to its degree $C=k$ 
with different bandwidth $B$ cases. 
The network parameter is $N=1000$, $m_0=m=3$.}
\end{figure}

In the second part, we investigate the effect of bandwidth on 
the network capacity considering the existence of different 
handling or delivering ability of nodes, i.e., in the case 
of $C$ is not a constant but propotional to the degree of 
each node $C=k$.
This may be used to descibe the fact that if a router is 
very important and bears heavy traffic, its 
delivering ability may be enhanced to avoid congestion. 

In the special case of $B \geq k_{max}$ corresponding to the 
maximum degree of the nodes in the network, the main 
difference from the case of $B \geq C=10$ is that 
the optimal value of local routing parameter $\alpha_c$ 
changes to $0.0$ while the maximum network capacity remains 
as $R_c \approx 40$ \cite{Wang}. 
This can be explained as follows. 
In the case of $B \geq k_{max}$ and $\alpha=0.0$, packets perform 
random-like walks on the network. 
A well-known result in the random walk process valid for this case 
is that the time the packet spends at a given node 
is proportional to the degree of such node in the limit of 
long times \cite{Bollob}.
One can easily conclude that, in the traffic system with 
many packets, the number of packets on a given 
node averaging over a period of time is proportional to 
the degree of that node, i.e., $n_i \sim k_i$. 
At the same time, the node delivering ability $C$ is 
proportional to its degree, i.e., $C_i \sim k_i$, 
so that the load and delivering ability of each node 
are balanced, which leads to a fact that no congestion 
occurs earlier on some nodes with particular degree 
than on others. 
Considering that in the traffic model, an occurrence of 
congestion at any node will diffuse to the entire network, 
no more easily congested nodes brings the 
maximum network capacity, so that routing packets with 
$\alpha=0.0$ can induce the maximum capacity of the system. 

Fig.\ref{Fig3} depicts the network capacity $R_c$ against 
$\alpha$ in the case of $C=k$ with different values of $B$.
One can see that the network capacity becomes smaller with 
$B$ decreasing, and the optimal value of $\alpha_c$ also 
decreases from $\alpha_c=0.0$ for $B=5$ to $\alpha_c=-0.1$ 
for $B=3$, $\alpha_c=-0.3$ for $1 \leq B \leq 5$, and 
$\alpha_c=-0.5$ for $B=1$.
The reason of capacity drop is the same as in the case of 
$C=10$, i.e., the bandwidth of the link prohibit the 
delivery process thus affect the network's overall capacity. 
The decrease of $\alpha_c$ is different from the case of 
$C=10$ and can be explained as follows. 
As mensioned before, $\alpha_c=0.0$ corresponds to 
$n_i(k) \sim k_i$ and $\alpha_c<0.0$ means redistributing 
traffic load in hub nodes to other noncentral nodes.
Considering the free flow condition of $n_{out} \geq n_{in}$
with the limitation of $n_{out}\approx C=k$, following a 
similar analysis, one can get
\begin{equation}\label{eq13}
\alpha \leq {\log {W\over B} \over \log k_i}.
\end{equation}

Or if considering $n_{out} \approx B k_i$, one can get 
\begin{equation}
\alpha \leq {\log W \over \log k_i}.
\end{equation}

In both cases, one can conclude that the optimal $\alpha$ should 
be close to zero. 
But when $\alpha=0.0$, $n_i \sim k_i$ and the nodes perform random 
selection among all its links to send packets. 
In the long run of $t \rightarrow \infty$, one can find that 
the number of packets forwarding towards each link in each 
time step should follow a Poisson distribution with mean value 
$\lambda=1$.
Thus the hub nodes are more easily jammed when $\alpha=0.0$ 
that some packets will be prohibit by the bandwidth of 
the links. 
Though the ideal condition is sending one packet per link in each 
time step, the bandwidth of the link should be more than $1$ to 
maintain free flow, i.e., $B=1+\delta$ where $\delta$ representing 
spanning of the Poisson distribution. 
Therefore, when $B$ is smaller than $1+\delta$, the optimal 
$\alpha_c$ should be smaller than zero to redistribute 
taffic load to other noncentral nodes in order to avoid 
congestion in hub nodes.
In Fig.\ref{Fig3}, one can see that when $B=5$, $\alpha_c$ remains 
as zero, whereas when $B$ decrease to less than $5$, $\alpha_c$ 
will decrease from zero.

This result is in agreement with the result of 
Yan et al.\cite{YanGang} and Wang et al.\cite{Wang2} 
that redistributing traffic load 
to the noncentral nodes can enhance the system's overall 
capacity.
The $\alpha_c$ smaller than zero can make the large degree nodes
fully used, and also allow packets to bypass those nodes when 
they afford heavy traffic burden. 
Thus the system's maximum efficiency is achieved.
Another interesting phenomenon emerges in the case of very low 
bandwidth is that the same optimal value of $\alpha_c=-0.5$ 
and $R_c\approx 8$ is obtained when $B=1$ in both cases of node 
capacity we considered in this paper, i.e., $C=10$ and $C=k$.
This simply show that the system's capacity is mainly controlled 
by the bandwidth of the links and the nodes' capacity do not 
affect the overall efficiency when bandwidth $B$ is very low.
That is, when the link bandwidth is low, it is useless 
to enhance the routers' ability only and traffic congestion 
would be triggered mainly by the links. 

\begin{figure}
\scalebox{0.9}[0.9]{\includegraphics{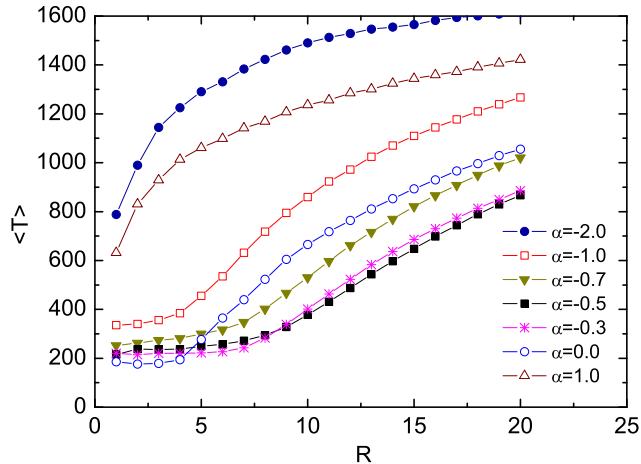}}
\caption{\label{Fig4}  (color online). 
The variation of packets average traveling time 
$\langle T \rangle$ versus $R$ with different value 
of $\alpha$ fixed. 
Other network parameters are $N=1000$, $m_0=m=3$,
$C_i=k_i$ and $B=1$.  }
\end{figure}

Fig.\ref{Fig4} shows the average travel time 
$\langle T \rangle$ against $R$ with different 
$\alpha$ when $C_i=k_i$.
The results are also in agreement with the above analysis 
that $\alpha_c=-0.5$ can lead to better efficiency of the 
network. 

\section{Summary}
In conclusion, we investigate the effects of link bandwidth 
on the traffic capability in scale-free network base on the 
local routing strategy. 
In general, the capacity decreases when the link bandwidth is 
considered, whether the node capacity is set as a constant or 
proportional to the degree of the nodes. 
Moreover, the optimal value of local routing paramter $\alpha_c$ 
also depends on the bandwidth of the links.
In the case of constant node capacity $C=10$, $\alpha_c$ increases 
from $-1.0$ to $-0.5$ when $B$ decreases from infinity to $1$, 
while in the case of $C=k$, $\alpha_c$ decreases from $0.0$ to 
$-0.5$ when $B$ decreases.
And we found that the node capacity can not enhance the system 
efficiency when $B$ is very low. 
We give analytical explanations for the above phenomena, and the 
analytical results are in agreement with the simulation results.
Moreover, we study the average traveling time of packets, which 
also exhibit similar phase transition behavior and is optimized 
when $\alpha$ is tuned to between $-1.0$ and zero.

Our study may be useful for evaluating the overall efficiency of 
networked traffic systems.
Moreover, this model can be applied to other networks and 
may also shed some light on alleviating the congestion of 
modern technological networks.

\section*{ACKNOWLEGEMENTS}
This work was funded by National Basic Research Program of China (No.2006CB705500),
the NNSFC under Key Project No.10532060, Project Nos. 70601026, 10672160, 
10404025, the CAS Dean Excellent Foundation, 
and by 
the Australian Research Council through a Discovery Project Grant.

\end{document}